\documentclass[12pt]{article}

\title{Conformal quantum mechanics  and Fick-Jacobs equation }
\author{Juan M. Romero\thanks{jromero@correo.cua.uam.mx}, 
Alejandro Gaona\thanks{2122800633@alumnos.cua.uam.mx}
\\[0.5cm]
\it Departamento de Matem\'aticas Aplicadas y Sistemas,\\
\it Universidad Aut\'onoma Metropolitana-Cuajimalpa\\
\it M\'exico, D.F  01120, M\'exico\\[0.3cm]} 

\begin{document}

\pagestyle{plain}

\maketitle

\begin{abstract}
It is found a relation between conformal quantum mechanics and Fick-Jacobs equation, which describes diffusion in channels.
This relation is given between a family of channels and a family of conformal Hamiltonians. In addition,  
 it is shown that a conformal Hamiltonian is associated with two channels with different geometry.
Furthermore exact solutions for Fick-Jacobs equation are  given  for this family of  channels.
\end{abstract}

\section{Introduction}
Recently,  mathematical techniques developed  in an area has been employed to study systems from other different areas.
In this subject, an amazing result is given by  $AdS_{d+1}/CFT_{d}$ duality, which  allows 
a relation between $(d+1)$-dimensional gravitational theory and certain classes of $d$-dimensional Yang-Mills theories \cite{maldacena:gnus}.  The conformal group  is very important in this duality, in fact this group 
is the  the largest  symmetry group of special relativity \cite{maldacena2:gnus}.  Now,  the  Schr\"odinger   group is a non-relativistic  conformal group 
\cite{hagen:gnus,hagen1:gnus}. This last   group   is the symmetry  group  for the free Schr\"odinger equation  and has been important  to study non-relativistic $AdS_{d+1}/CFT_{d}$ duality \cite{son:gnus,son1:gnus}.  To study $AdS_{2}/CFT_{1}$ correspondence, the so call   conformal quantum mechanics has been proposed as $CFT_{1}$  dual to $AdS_{2},$ see  \cite{jackiw:gnus,jackiw1:gnus}. The  conformal quantum mechanics is invariant under Schr\"odinger group and 
has been employed to  study   problems  from black-holes to atomic physics \cite{jackiw2:gnus,jackiw3:gnus,jackiw4:gnus}. 
Furthermore, the simplest model of diffusion is described by the Fick equation  and Sophus Lie  showed that this equation is invariant under   Schr\"odinger group \cite{lie:gnus}. Other studies about   diffusion phenomena and  Schr\"odinger group can be seen in \cite{henkel:gnus}.  
Then, the conformal symmetry, relativistic or non-relativistic, is very important to understand diverse aspects of different systems. \\

Now, when  the diffusion is in a  channel, which has the shape of  surface of revolution with cross sectional  area $A(x),$  the Fick equation has to be  changed to  Fick-Jacobs  equation \cite{jacobs:gnus}
\begin{eqnarray}
\frac{ \partial C(x,t)}{\partial t}&=&\frac{\partial }{\partial x}
\left[D_{0}A(x) \frac{\partial }{\partial x} \left(
\frac{C(x,t)}{A(x)} \right)\right],\label{eq:fick0} 
\end{eqnarray}
where $C(x,t)$ is the particle concentration  and $D_{0}$ is the diffusion coefficient.
This last equation  is important to study diffusion in biological channels or zeolites 
\cite{reguera:gnus,reguera1:gnus,reguera2:gnus,dagdug:gnus,dagdug1:gnus,zwanzig:gnus,zwanzig1:gnus}.  The Fick-Jacobs equation does not look like the free Schr\"odinger equation, but it can be mapped to Schr\"odinger equation with an effective potential \cite{oswal:gnus}. \\

In this paper we will show that the  Fick-Jacobs equation is equivalent to conformal quantum mechanics for a family  of  channels. Then for this family of channels 
 the  Fick-Jacobs equation is  invariant under  Schr\"odinger group. Also, it is found that the equivalence  is given between a family of channels and a family of conformal Hamiltonians.  In addition,  it is shown that a conformal Hamiltonian is associated with two channels with different geometry. 
 For these channels  an exact solution for the  Fick-Jacobs equation  is given.\\

This paper is organized in the following way: in section $2$ a brief review about Schr\"odinger group and conformal quantum mechanics is given; in section $3$ it is shown   
that the  Fick-Jacobs equation is equivalent to conformal quantum mechanics for a set of  particular channels and an exact solution for this  equation  is given. Finally, in section $4$ a summary is given.

\section{Schr\"odinger group}

The free Schr\"odinger equation   
\begin{eqnarray}
 i\hbar \frac{\partial\psi\left(\vec x,t\right)}{\partial t} =
 -\frac{\hbar^{2}}{2m} \nabla ^{2} \psi\left(\vec x,t\right), 
\end{eqnarray}
is invariant under the  following transformation: Galileo transformation  $ x_{i}^{\prime}=x_{i}+v_{i}t,$ rotations $x_{i}^{\prime}=R_{ij}x_{j},$
 space-time translation $t^{\prime}=t+a, x_{i}^{\prime}=x_{i}+x_{0i},$
 anisotropic scaling  $t^{\prime}=b^{2}t ,\quad  x_{i}^{\prime}=b x_{i}$  and special conformal transformation \cite{hagen:gnus,hagen1:gnus}
\begin{eqnarray}
t^{\prime}=\frac{t}{1+at },\qquad x_{i}^{\prime}=\frac{x_{i} }{1+at }.\label{eq:sct}
\end{eqnarray}
Some work about Schr\"odinger group and conformal symmetry can be seen in \cite{ph1:gnus,ph2:gnus,ph3:gnus,ph4:gnus, ph5:gnus,ph6:gnus,ph7:gnus,ph8:gnus}.\\

Now,  the Schr\"odinger equation for  the $1$-dimensional  conformal quantum mechanics  is given by 
\begin{eqnarray}
i\hbar \frac{\partial \psi(x,t)}{\partial t} =H\psi(x,t),\qquad H=-\frac{\hbar^{2}}{2m} \frac{\partial^{2} }{\partial x^{2}} +\frac{g}{x^{2}}, \label{eq:cqmh}
\end{eqnarray}
which is invariant under Schr\"odinger transformation. The classical system with the potential $V(r)=gr^{-2}$ was first studied by Jacobi \cite{jacobi:gnus} and  the quantum system was   proposed by Jackiw \cite{jackiw5:gnus}. Using the Schr\"odinger group generators, the spectrum of  Hamiltonian (\ref{eq:cqmh}) was  found  by  de Alfaro,  Fubini and  Furlan  \cite{jackiw6:gnus}.  The Hamiltonian (\ref{eq:cqmh}) appears in different contexts, from black-holes to atomic physics \cite{jackiw2:gnus,jackiw3:gnus,jackiw4:gnus}. In the next section we will show that this systems also appears in diffusion phenomena.

\section{Conformal quantum mechanics and Fick-Jacobs equation}

Using $C(x,t)=\sqrt{A(x)}\psi(x,t),$  the Fick-Jacobs equation becomes 
\begin{eqnarray}
\frac{ \partial \psi(x,t)}{\partial t}=\left[D_{0} \frac{\partial^{2} }{\partial x^{2}} 
-\frac{D_{0}}{2\sqrt{A(x)}}
\frac{\partial}{\partial x}\left(\frac{1}{\sqrt{A(x)}} \frac{\partial A(x)}{\partial x}\right)\right]\psi(x,t).
\end{eqnarray}
Then, if we propose $\psi(x,t)=e^{-Et}\phi(x),$ we get the following Schr\"odinger equation
\begin{eqnarray}
E\phi(x)=H\phi(x),\label{eq:static}
\end{eqnarray}
where
\begin{eqnarray}
H=-D_{0} \frac{\partial^{2} }{\partial x^{2}} +
\frac{D_{0}}{2\sqrt{A(x)}} \frac{\partial}{\partial x}\left(\frac{1}{\sqrt{A(x)}} \frac{\partial A(x)}{\partial x}\right).
\label{eq:fjhc}
\end{eqnarray}
Now, the family of channels with  cross sectional area $A(x)=ax^{2\nu}$ is associated  with  the following  family of Hamiltonians
\begin{eqnarray}
H=-D_{0} \frac{\partial^{2} }{\partial x^{2}} +\frac{g}{x^{2}}, \qquad g=D_{0}\nu\left(\nu-1\right).  
\label{eq:fjh}
\end{eqnarray}
 For each $\nu$ we have a conformal quantum mechanics Hamiltonian (\ref{eq:cqmh}). However, for each Hamiltonian  (\ref{eq:fjh}) we have 
two channels, namely  each  Hamiltonian is associated with two $\nu$ values. For example, 
 $\nu=0$ and  $\nu=1,$ represent different sectional areas, but both cases  give the same  Hamiltonian
\begin{eqnarray}
H=-D_{0} \frac{\partial^{2} }{\partial x^{2}}. 
\end{eqnarray}

The solution for the Schr\"odinger equation (\ref{eq:static}) with  the Hamiltonian (\ref{eq:fjh}) is given by
\begin{eqnarray}
\phi_{\nu}(x)=|x|^{\frac{1}{2}}J_{\pm\left(\frac{2\nu-1}{2}\right)} \left( \pm \sqrt{\frac{E}{D_{0}}} x\right),
\end{eqnarray}
where $J_{p} \left( w\right)$ is the Bessel function of  order $p.$ Then, if  the channel has  cross sectional area $A(x)=ax^{2\nu},$
the solution for the Fick-Jacobs  is given by 
\begin{eqnarray}
C_{\nu}(x,t)=Be^{-Et}|x|^{\frac{2\nu+1}{2}} J_{\pm\left(\frac{2\nu-1}{2}\right)} \left( \pm \sqrt{\frac{E}{D_{0}}} x\right),
\end{eqnarray}
here $B$ is a constant.\\

Notice that whether  $\nu=0$ the solution  
\begin{eqnarray}
C_{\nu=0}(x,t)=e^{-Et}\left(  B_{1}\sin \left( \sqrt{\frac{E}{D_{0}}} x\right)+B_{2}\cos \left( \sqrt{\frac{E}{D_{0}}} x\right) \right),
\end{eqnarray}
is obtained. While if   $\nu=1,$ the solution
\begin{eqnarray}
C_{\nu=1}(x,t)=e^{-Et} |x| \left(  B_{1}\sin \left( \sqrt{\frac{E}{D_{0}}} x\right)+B_{2}\cos \left( \sqrt{\frac{E}{D_{0}}} x\right) \right)
\end{eqnarray}
is gotten. We can see that $\nu=0$ and $\nu=1$ are associated with the same Hamiltonian, but  the particle concentration is not the same.

\section{Summary}
In this paper we shown a relation between conformal quantum mechanics and Fick-Jacobs equation.
This relation is given between a family of channels and a family of conformal Hamiltonians. It was found  
 that a conformal Hamiltonian is associated with two channels with different geometry. In addition, 
 exact solutions for Fick-Jacobs equation are  given  for this family of  channels. 
This result is interesting, because the conformal quantum mechanics has been proposed as a realization of  $AdS_{2}/CFT_{1}$ duality 
and  Fick-Jacobs equation is employed to describe diffusion in biological channel.  Then, it is possible that mathematical 
techniques from string theory  can be employed to study some biological problems.  

\section*{Acknowledgments}
We would like  to thank  P.A Horvathy for his comments about the history of Schr\"odinger symmetry.


\begin{thebibliography}{99}


\bibitem{maldacena:gnus}
J. Maldacena, {\it The Large N limit of Superconformal Field Theories and Supergravity,}  Adv Theor. Math. Phys. 2, 231 (1998), [hep-th/9711200].


\bibitem{maldacena2:gnus}
O. Aharony,  S. S. Gubser, J.  Maldacena, H.  Ooguri, {\it Large N field theories, string theory and gravity,}
Phys.Rept. 323 (2000) 183-386,  [hep-th/9905111].


\bibitem{hagen:gnus}
C. R. Hagen, {\it Scale and conformal transformations in galilean-covariant field theory,}  Phys. Rev. D {\bf 5}  377 (1972).

\bibitem{hagen1:gnus} 
U. Niederer, {\it The maximal kinematical invari- ance group of the free Schrodinger equation,} Helv. Phys. Acta {\bf 45} (1972) 802.


\bibitem{son:gnus}  D. T. Son, {\it Toward an AdS/cold atoms correspondence: a geometric realization of the Schroedinger symmetry,} 
Phys. Rev. D {\bf 78}, 046003 (2008),  [arXiv:0804.3972 [hep- th]].

\bibitem{son1:gnus} 
 K. Balasubramanian and J. McGreevy, {\it Gravity duals for non-relativistic CFTs,} Phys. Rev. Lett. 101, 061601 (2008), [arXiv:0804.4053 [hep-th]].

\bibitem{jackiw:gnus}
C. Chamon, R. Jackiw, S. -Y. Pi and L. Santos, {\it Conformal quantum mechanics as the CFT1 dual to AdS2,} Phys. Lett. B {\bf 701}, 503 (2011),  [arXiv:1106.0726 [hep-th]]. 


\bibitem{jackiw1:gnus}
R. Jackiw, S.-Y. Pi, {\it Conformal Blocks for the 4-Point Function in Conformal Quantum Mechanics,}
 Phys.Rev. D {\bf 86}, (2012) 045017, Erratum-ibid. D86 (2012) 089905,  [arXiv:1205.0443 [hep-th]] 
 
\bibitem{jackiw2:gnus}
R. Britto-Pacumio, J. Michelson, A. Strominger, A.
Volovich, {\it Lectures on superconformal quantum mechanics and multiblack hole moduli spaces,}
 Contributed to NATO Advanced Study Institute on Quantum Geometry. 
Published in Progress in string theory and M-theory, 235-264 (1999), [hep-th/9911066].

\bibitem{jackiw3:gnus}
H. E. Camblong, C. R. Ordonez, {\it Anomaly in conformal quantum mechanics: From molecular physics to black holes,} Phys. Rev. D {\bf 68}, 
125013 (2003), [hep-th/0303166].


\bibitem{jackiw4:gnus}
K. Sakamoto, K. Shiraishi, {\it  Conformal Quantum Mechanics in Two Black Hole Moduli Space,} 
Phys.Rev. D {\bf 66} (2002) 024004, [arXiv:hep-th/0203152].

\bibitem{jackiw5:gnus}
R. Jackiw, {\it Introducing Scale Symmetry,} Phys. Today {\bf 25} (1), 23 (1972).

\bibitem{jackiw6:gnus}
V. de Alfaro, S.Fubini and G. Furlan,  {\it Conformal Invariance in Quantum Mechanics,}  Nuovo Cim. {\bf 34} A, 569 (1976).


\bibitem{lie:gnus}
S. Lie, Arch. Math. Nat. vid. (Kristiania) 6, 328 (1882).

\bibitem{henkel:gnus}
S. Stoimenov, M. Henkel, {\it Dynamical symmetries of semi-linear Schršdinger and diffusion equations,}
Nucl. Phys. B723, 205-233 (2005), 	[arXiv:math-ph/0504028].

\bibitem{jackiw6:gnus}
V. de Alfaro, S.Fubini and G. Furlan,  {\it Conformal Invariance in Quantum Mechanics,}  Nuovo Cim. {\bf 34} A, 569 (1976).

\bibitem{jacobs:gnus}
M. Jacobs, {\it Diffusion Processes,} Springer, New York, (1967).


\bibitem{reguera:gnus}
D. Reguera, J. M. Rub\'i, {\it Kinetic equations for diffusion in the presence of entropic barriers,}
Phys. Rev. E {\bf 64}, 061106 (2001).

\bibitem{reguera1:gnus}
D. Reguera, G. Schmid, P. S. Burada, J. M. Rub\'i, P. Reimann, P.
H\"{a}nggi {\it Entropic Transport: Kinetics, Scaling, and Control
Mechanisms,} Phys. Rev. Lett. {\bf 96}, 130603 (2006).

\bibitem{reguera2:gnus}
P. S. Burada, P. H\"{a}nggi, F. Marchesoni, G. Schmid, P. Talkner,
{\it Diffusion in Confined Geometries,} Chem. Phys. Chem. {\bf 10}, 45
(2009).


\bibitem{dagdug:gnus}
A. M. Berezhkovskii, M. A. Pustovoit, S. M. Bezrukov, {\it Diffusion
in a tube of varying cross section: Numerical study of reduction to
effective one-dimensional description,} J. Chem. Phys. {\bf 126}, 134706
(2007).

\bibitem{dagdug1:gnus}
 I. Pineda, M. V. V\'azquez, L. Dagdug, {\it Diffusion
between Two Chambers Connected by a Conical Capillary,} in {\it New
trends in statistical physics} Edited by A. Macias, L. Dagdug,
(World Scientific, Singapore 2010) pp. 147-162.

\bibitem{zwanzig:gnus}
R. Zwanzig,  {\it Diffusion past an entropy barrier,} {\bf 96}, J. Chem. Phys.3926 (1992).

\bibitem{zwanzig1:gnus}
P. Kalinay,  J. K. Percus, {\it Corrections to the Fick-Jacobs equation,}
Phys. Rev. E {\bf 74}, 041203 (2006).

\bibitem{oswal:gnus}
J. M. Romero,  O. Gonz\'alez-Gaxiola,  G. Chac\'on-Acosta,  {\it Exact solution to Fick-Jacobs equatio,}
 to appears in  International Journal of Pure and Applied mathematics (IJPAM), Volume 82, No. 1 (2013), [arXiv:1108.1164 [math-ph]] .


\bibitem{ph1:gnus} 
C. Duval,  P. A. Horvathy, {\it Non-relativistic conformal symmetries and Newton-Cartan structures,}
 J. Phys. A {\bf  42}, 465206 (2009),  [arXiv:0904.0531 [math-ph]].
 
\bibitem{ph2:gnus} 
C. Duval, G. W. Gibbons, P. Horvathy, 
{\it Celestial mechanics, conformal structures and gravitational waves,}
 Phys. Rev. D {\bf 43}  3907-3922 (1991), [arXiv:hep-th/0512188].

\bibitem{ph3:gnus}  
 C. Duval, M. Hassaine,  P. A. Horvathy,
{\it The geometry of Schr\"odinger symmetry in non-relativistic CFT,}
Ann. Phys. {\bf 324} 1158  (2009), [arXiv:0809.3128 [hep-th]].

\bibitem{ph4:gnus}  
C.  Leiva, M. S. Plyushchay, {\it Conformal symmetry of relativistic and nonrelativistic systems and 
$Ads/ CFT$ correspondence,}
Annals Phys. {\bf 307}  (2003) 372-391, [arXiv:hep-th/0301244].

\bibitem{ph5:gnus}  
C. Leiva, M. S. Plyushchay, {\it Superconformal mechanics and nonlinear supersymmetry,}
JHEP {\bf 0310}  069 (2003), [arXiv:hep-th/0304257].
 
\bibitem{ph6:gnus}
A.  Anabalon, M. S. Plyushchay, {\it Interaction via reduction and nonlinear superconformal symmetry,}
Phys.Lett. B {\bf 572} 202-209 (2003), [arXiv:hep-th/0306210].

\bibitem{ph7:gnus}
A. Galajinsky, {\it Remark on quantum mechanics with conformal Galilean symmetry,}
 Phys. Rev.  D {\bf 78}  087701 (2008), [arXiv:0808.1553 [hep-th]]. 
 
 \bibitem{ph8:gnus}
 A. Galajinsky, I.V. Masterov, {\it Remark on quantum mechanics with N=2 Schr\"odinger supersymmetry,}  
 Phys. Lett. B {\bf 675}  116  (2009), [arXiv:0902.2910 [hep-th]].
 
 \bibitem{jacobi:gnus}
 C. G. J. Jacobi, {\it Vorlesungen uber Dynamik,} Univ. K\"onigsberg 1842-43. 
 
\end{thebibliography}
\end{document}